\documentclass{Rinton-P9x6}
\usepackage{amsmath,amscd,amssymb}

\begin{document}

\title{Dirichlet Spheres in Continuum Quantum Field 
Theory\footnote{T\lowercase{alk presented at at the} QFEXT03 
w\lowercase{orkshop}, N\lowercase{orman}, 
O\lowercase{klahoma}, S\lowercase{eptember} 2003.}}

\author{H. Weigel}

\address{Fachbereich Physik, Universit\"at Siegen,\\
Walter--Flex--Stra{\ss}e 3, D--57068 Siegen \\
E-mail: weigel@physik.uni-siegen.de}


\maketitle

\abstracts{We study the vacuum polarization (Casimir) energy in 
renormalizable, continuum quantum field theory in the presence of 
a background field, designed to impose Dirichlet boundary conditions
on the fluctuating quantum field.  
In two and three spatial dimensions the Casimir energy diverges 
as a background field becomes concentrated on the surface on which 
the Dirichlet boundary condition would eventually hold. This divergence 
does not affect the force between rigid bodies, but it does invalidate 
calculations of Casimir stresses based on idealized boundary conditions.}

\section{Introduction and Motivation}

This presentation is based on work done in collaboration\cite{Gr03} 
with N.~Graham, R. L.~Jaffe, V.~Khemani, M.~Quandt, and 
O. Schr\"oder.

The interaction of fluctuating quantum fields with background 
fields gives rise to vacuum polarization energies because the 
zero--point energies change. Colloquially these vacuum polarization 
energies are called Casimir energies. In this talk I present 
calculations\cite{Gr03} of Casimir energies for extreme cases when a 
static background field in $D$ spatial dimensions, $\sigma(\vec{x\,})$, 
imposes Dirichlet boundary conditions on the fluctuating quantum field, 
$\phi(t,\vec{x\,})$, in some subspace $\mathbb{S}\subset \mathbb{R}^D$: 
$\phi(t,\vec{a\,})=0$ for $\vec{a\,}\in \mathbb{S}$. These calculations 
are afflicted with severe divergences which fall into two categories: 
(i) ultraviolet divergences in quantum field theory and (ii) those due 
to singularities of the background fields. Although the vacuum 
polarization energy cannot be reliably computed from only the low order 
Feynman diagrams, the type (i) divergences can be identified thereof. 
The type (ii) divergences are due to large Fourier components carried 
by the background fields. In reality the contributions of these components 
to the vacuum polarization energy are suppressed 
by the dynamics of the interaction of some material with the quantum 
field. We therefore label these divergences {\em quantum} and 
{\em material} divergences for type (i) and (ii), respectively.
It is essential to disentangle the quantum and material divergences
since no reliable conclusion can be drawn from a calculation
that does not have the material divergences under control.

\enlargethispage{0.3cm}
As is reviewed in Ref.\cite{Leipzig} the existence of scattering
data for the interaction of the quantum field with the background
allows us to unambiguously compute the corresponding vacuum 
polarization energy\footnote{See also the contributions by
R.L. Jaffe and M. Quandt to these proceedings. A detailed derivation
is given in Ref.\cite{Gr02a}.}. In particular,
the associated quantum divergences are under control when the
theory that describes this interaction is renormalizable. Hence we 
can compute the Casimir energy for any (piecewise) continuous
background field. The strategy to explore the material divergences
is thus to consider a background field that in a particular, singular 
limit imitates a Dirichlet boundary condition. Most importantly,
the vacuum polarization energy is computed {\em before} that
singular limit is assumed. We then study the vacuum polarization
energy as a function of the parameters that regulate this singular 
limit. To be specific, the Dirichlet boundary condition at a 
point $\vec{a\,}$ is imitated by a delta--function background,
$\sigma(\vec{x\,})=\lambda \delta(\vec{x\,}-\vec{a\,})$ in the
{\em strong} limit, $\lambda\to\infty$. We parameterize the 
delta--function with the help of a finite width, $\Delta$, via
\begin{eqnarray}
    \sigma_{\vert}(z)=\frac{\lambda}{\Delta}
\Big(\theta(z+\Delta/2)-\theta(z-\Delta/2)\Big)\cr
    \sigma_{\parallel}(z)=\frac{\lambda}{\Delta}
       \Big(\theta (|z|-L+\Delta/2 )
        -\theta (|z|-L-\Delta/2) \Big)\cr
    \sigma_{\circ}(r)=
        \frac{3\lambda/4\pi}{(R+\Delta)^3-R^3}
        \Big(\theta(r-R)-\theta(r-R-\Delta)\Big)
\label{background}
\end{eqnarray}
for the cases of a single plate, two plates at distance $2L$, and
a sphere of radius $R$ in three spatial dimensions, respectively.
For plates the total number of spatial dimensions is of lesser
importance as we only consider the energy per unit area. The above
slab--type parameterizations of the delta--functions in various
geometries are advantageous because they have simple Fourier transforms,
\begin{equation}
\tilde\sigma(\vec{p\,})=\int d^Dx \, {\rm e}^{i\vec{p}\cdot\vec{x}}\,
\sigma(\vec{x\,})
\label{Ftrans}
\end{equation}
for all spatial dimensions $D$. For any $\lambda<\infty$ and 
$\Delta>0$ the vacuum polarization energy can unambiguously computed
with the techniques of Refs.~\cite{Leipzig,Gr02a}.
The {\em sharp} limit ($\Delta\to0$) imitates the delta--function 
background. The subsequent {\em strong} limit imposes the Dirichlet 
condition. This calculation is different from a pure boundary condition 
calculation which assumes both the sharp and the strong limits {\em before}
computing the vacuum polarization energy\cite{Miltonrep}.

As motivated above, we study the low order Feynman diagrams to analyze the
emergence of material divergences in the sharp limit. For this we consider
a bosonic quantum field whose interaction with the static background field
is given by the Lagrangian 
${\cal L}_{\rm int}=\frac{-1}{2}\sigma(\vec{x\,})\phi^2$.

\section{The Tadpole Graph}

Let us start the discussion of low order contribution to the
vacuum polarization energy by considering the tadpole diagram 
\begin{equation}
\begin{minipage}{2.6cm}\epsfig{file=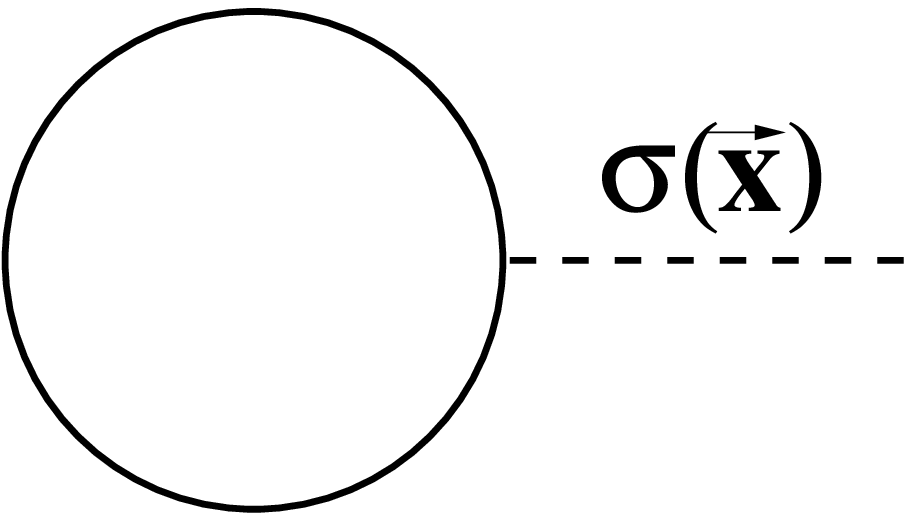,width=2.4cm,height=1cm}
\end{minipage}
=E^{(1)}=\int d^Dx\, \sigma(\vec{x\,})
\int \frac{d^n l}{(2\pi)^n}\,
\frac{i}{l^2-m^2}
\label{tadpole}
\end{equation}
which we have written in dimensional regularization, {\it e.g.}
$n$ may be fractional while $D$ is an integer. It is actually important,
not to include the background into dimensional regularization because
renormalization occurs on the level of Green's functions which do not
involve the background fields. We recognize that spatial and
loop integration factorize, {\it i.e.} the tadpole diagram is local. 
Within the non--tadpole renormalization condition the counterterm 
contribution therefore exactly cancels $E^{(1)}$ regardless of
the details of $\sigma(\vec{x\,})$. Thus it is obvious that there are 
no further divergences at this order.

\section{The Two--Point--Function}

We will now study the two--point--function for which 
analytical results are also available. There is no quantum 
divergence at this order in two spatial dimensions and the
diagram can straightforwardly be computed 
\begin{equation}
\begin{minipage}{2.6cm}\epsfig{file=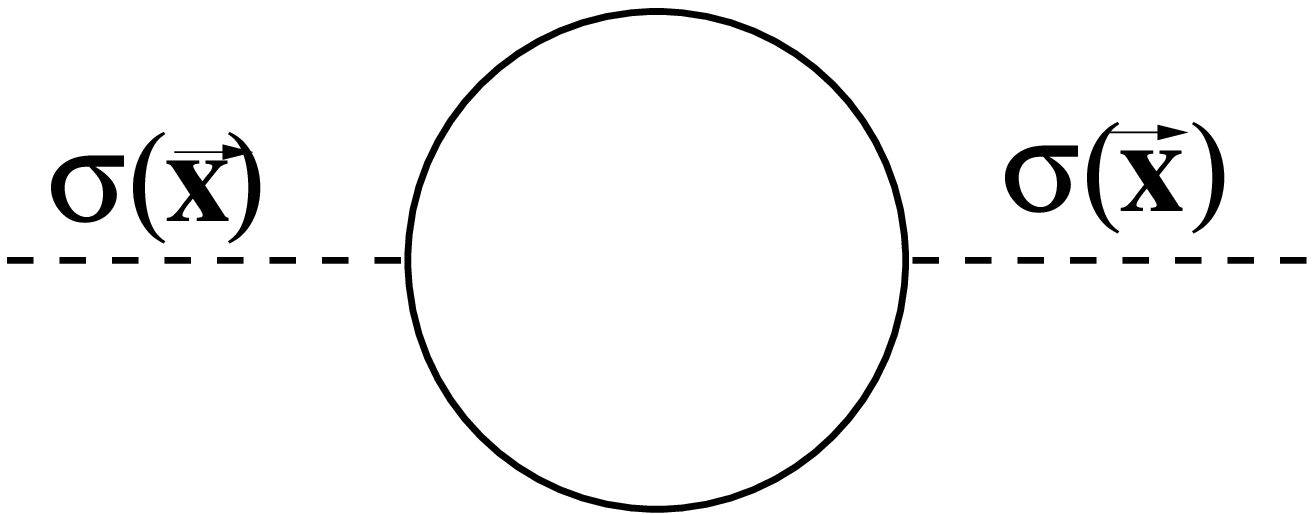,width=2.4cm,height=1cm}
\end{minipage}
=E_{\rm 2D}^{(2)}[\sigma] =-\frac{\lambda^2}{16\pi}
\int \frac{d^{\,2}p}{(2\pi)^2}\,
\tilde{\sigma}(\vec{p\,})\tilde{\sigma}(-\vec{p\,})\,
\frac{1}{|\vec{p\,}|}{\rm arctan}\frac{|\vec{p\,}|}{2m}\,.
\label{twopt2d}
\end{equation}

It is then straightforward to extract the most singular 
contribution to the vacuum polarization energy per unit
length 
\begin{equation}
{\cal E}^{(2)}_{\rm 2D}(\sigma_{\vert}) =  \frac{1}{32 \pi}
\ln{(\Delta m)} + \textrm{terms~finite~for~} \Delta \to 0
\label{e2d2}
\end{equation}
for a single plate, {\it cf.} eq.~(\ref{background}). Obviously
this contribution diverges logarithmically in the sharp limit.
A quadratic counterterm, that would anyhow not be required by
any quantum divergence, diverges like $1/\Delta$ and thus does
not remove this divergence. 

In three spatial dimensions the two--point--function has a 
quantum divergence which is manifest in the polarization 
tensor 
\begin{equation}
\begin{minipage}{2.6cm}\epsfig{file=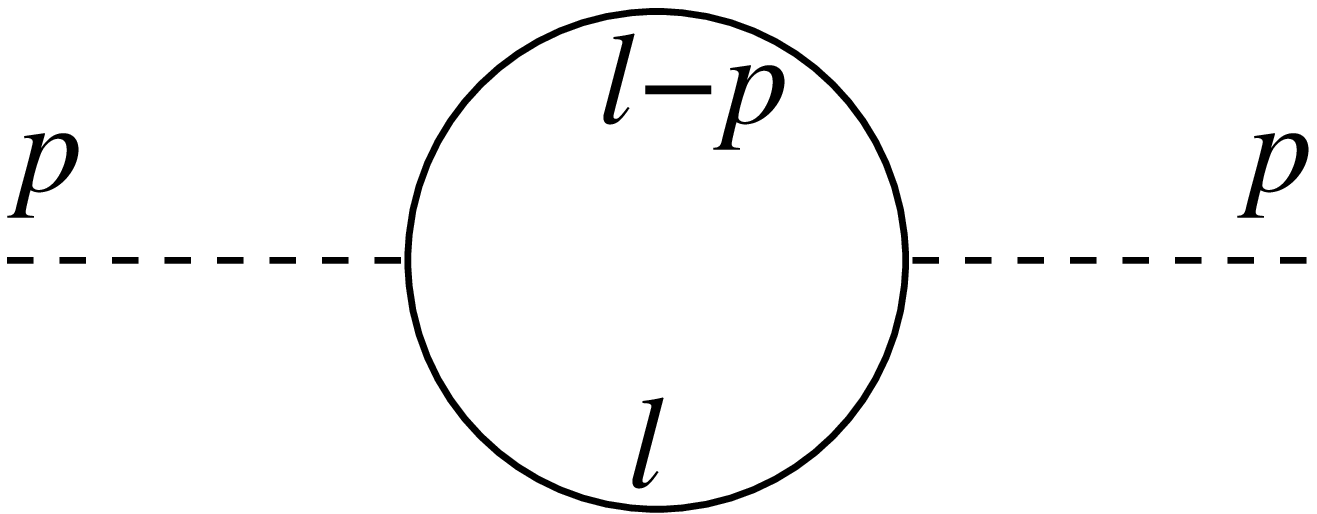,width=2.4cm,height=1cm}
\end{minipage}
=\Pi(p^2)=\int \frac{d^nl}{(2\pi)^n}\,
\frac{1}{l^2-m^2}\,\frac{1}{(l-p)^2-m^2}\,.
\label{poltensor}
\end{equation}
The associated divergences are removed by a quadratic 
counterterm ${\cal L}_{\rm CT}\propto\Pi(-\mu^2)\sigma^2$, 
whose coefficient is entirely given by the polarization 
tensor and thus is the {\em same} for all background fields.
This independence is an essential feature of renormalization:
counterterm coefficients are fixed by a (chosen) condition but 
not modified when the background changes. Then the second order
contribution to the renormalized vacuum polarization energy is 
then easily computed by contracting the renormalized vacuum 
polarization tensor with the Fourier transform of the background
field. To be concise, we only display the result
for the case that the fluctuating field is massless
\begin{equation}
E_{\rm 3D}^{(2)}[\sigma] =\frac{1}{64\pi^2} \int
\frac{d^{\,3}p}{(2 \pi)^3}\,
\tilde{\sigma}(\vec{p\,}) \,\tilde{\sigma}(-\vec{p\,})\,
{\ln}\frac{\vec{p\,}^2}{\mu^2}\,.
\label{e2d3}
\end{equation}

For the geometry of two parallel plates with separation $2L$ this yields
\begin{equation}
{\cal E}^{(2)}_{\rm  3D}[\sigma_{\parallel}] =
-\frac{\lambda^2}{16\pi^2 \Delta} \left\{\ln{(\Delta \mu)}
+ \gamma-1 +\frac{\Delta}{4L} +
{\cal O}\left( \frac{\Delta^{2}}{L^{2}}\right)\right\}
\label{e2plates}
\end{equation}
for the energy ${\cal E}=E/A$ per unit area $A$. We observe a 
$\ln{\Delta}\,/\Delta$ singularity in the sharp limit. This
singularity, however, is just twice the one for a single plate
and it does not depend on the separation of the two plates. Hence
the force (density) $\partial{\cal E} /\partial L$ does not diverge
in the sharp limit. 

The situation is different for the sphere,
\begin{eqnarray}
E^{(2)}_{\rm 3D}[\sigma_{\circ}]&=&\frac{-\lambda^2}{128\pi^3 R^2}
\left(\frac{1}{\Delta} \left(\ln{(\mu\Delta)}+\gamma-1\right)-
\frac{\ln{(\mu\Delta)}}{R} \right) + \ldots \,,
\label{e2sphere}
\end{eqnarray}
where the ellipses represent contributions that are finite as 
$\Delta\to0$. Again we observe a $\ln{\Delta}\,/\Delta$ singularity 
in the sharp limit. This time, however, the singularity depends on
the radius of the sphere. Thus the stress $\partial{\cal E}/\partial R$
does in fact diverge and remains ill--defined within a pure boundary 
condition calculation; the width $\Delta$ cannot approach zero. Rather 
it is bounded by some material properties.

\section{The Three--Point--Function}

\begin{figure}
\centerline{\epsfig{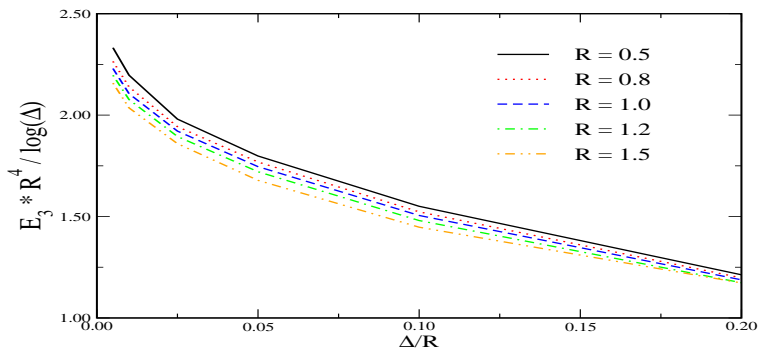}}
\caption{\label{fig1}The three--point--function contribution to
the vacuum polarization energy of a spherical shell, 
$E^{(3)}[\sigma_\circ]$ as a function 
of the inner radius ($R$) and the width ($\Delta$) of the
shell, {\it cf.} eq.~(\protect\ref{background}).
Factors that are independent of $L$ or $\Delta$ have been omitted.
Dimensions are set by the mass of the fluctuating quantum field.}
\end{figure}

In case of the three--point--function 
\begin{equation}
E^{(2)}[\sigma]=\,
\begin{minipage}{2.5cm}\epsfig{file=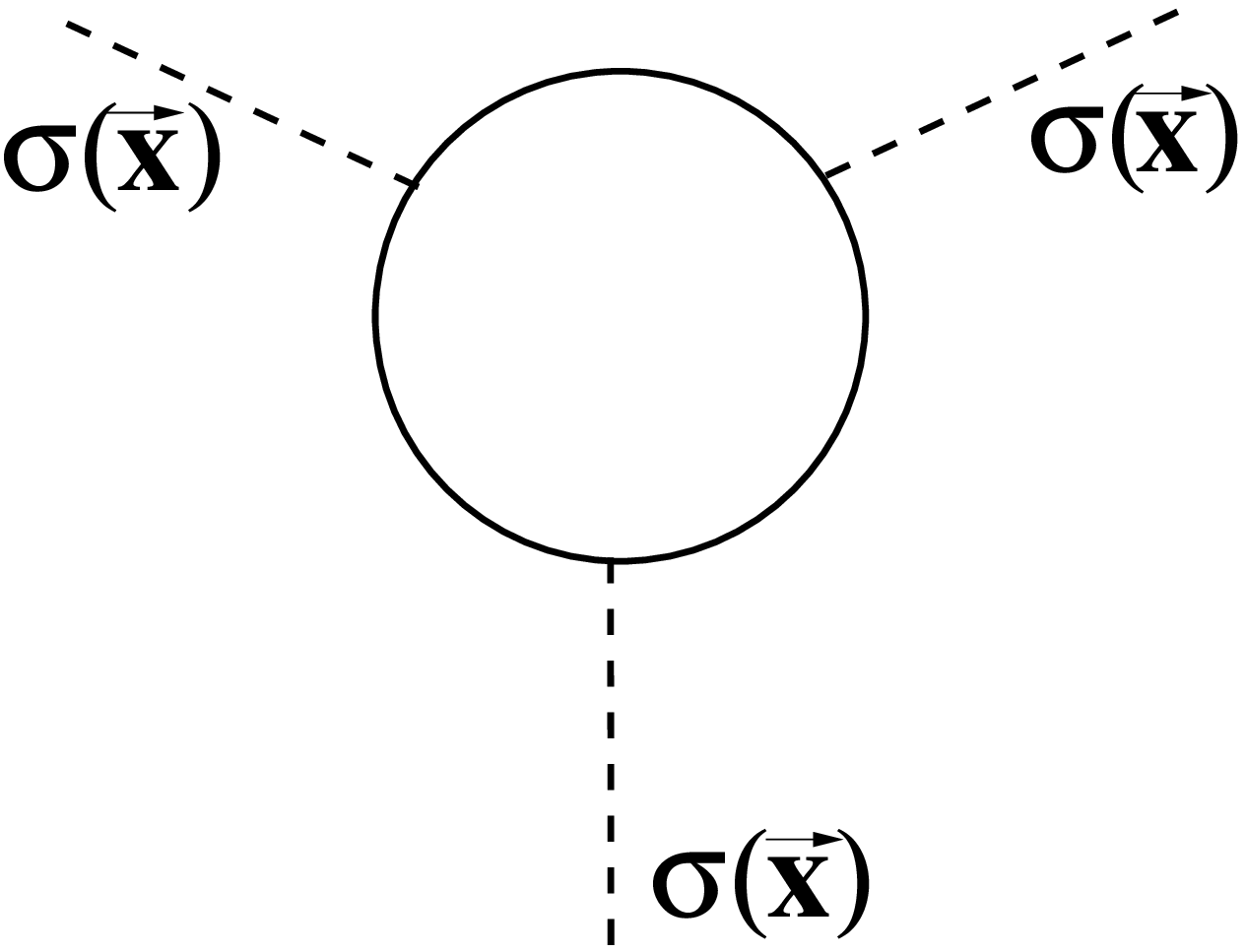,width=2.3cm,height=1.2cm}
\end{minipage}
\label{threeptfct}
\end{equation}
only numerical results are available. For the parallel plates geometry
they are listed in table~\ref{tab1}. Essentially we observe a 
logarithmic divergence in the sharp limit. Here the divergence is less
severe than for the two point function which indicates that the 
higher order contributions are actually finite. This will be discussed
in section 5. The important result is that this divergence is 
independent of the separation of the two plates implying that the
force has a finite sharp limit.
\begin{table}
\caption{\label{tab1}The three point function contribution to the 
energy density for two parallel plates ${\cal E}^{(3)}[\sigma_\parallel]$
as a function of their separation~($2L$) and the width of the 
slabs~($\Delta$), {\it cf.} eq.~(\protect\ref{background}). 
Factors that are independent of $L$ or $\Delta$ have been omitted.
Dimensions are set by the mass of the fluctuating quantum field.}
\centerline{
\begin{tabular}{l|c c c c c}
$\Delta$ & $L=0.5$ & $L=0.8$ & $L=1.0$ & $L=1.2$ & $L=1.5$\\
\hline
0.010 & -163.04 & -160.24 & -160.88 & -161.04 & -161.28 \\
0.025 & -122.16 & -119.68 & -119.68 & -119.76 & -120.00 \\
0.050 & -95.36 & -92.64 & -92.16 & -92.08 & -92.24 \\
0.100 & -70.56 & -67.84 & -67.36 & -67.28 & -67.20 \\
0.200 & -48.08 & -44.72 & -45.12 & -45.04 & -44.96 \\
\end{tabular}}
\end{table}
Also for the spherical shell we find a logarithmic singularity
as the sharp limit is approached. This is shown in figure~\ref{fig1}.
In this case, however, the divergent term depends on the radius
of the shell and thus the resulting stress, $\partial E/\partial R$
is also divergent, as already observed for the two--point--function
contribution to the vacuum polarization energy. One might want to
argue that a suitable constant of proportionality in the definition of 
the spherical background field, $\sigma_\circ$, in eq.~(\ref{background})
might yield a finite force. However, this cannot be achieved for
both the two-- and three--point--function contributions\cite{Gr03}.

The results obtained so far generalize: The renormalized vacuum
energies always diverge in the boundary condition limit while forces
between rigid bodies have a finite sharp limit because the singularities
are (material) properties of the individual objects. However, for stresses 
that arise from deformations of such objects the singularities do not 
drop out.

\section{The Sharp Limit at $\mathbf{n}^{\rm th}$ order}

In this section we will take a different venue and adopt the
sharp limit first and compute Feynman diagrams for the vacuum
polarization energy. These diagrams are potentially divergent
and we regularize with a sharp cutoff $\Lambda$. As indicated
in the previous section this shows that diagrams of large
enough order indeed do not diverge in the sharp limit. Note, 
however, that this procedure does not make contact with 
renormalization theory, in fact, we do not renormalize at all.
Rather it is an efficient way to localize the divergences for
delta--function type background potentials. For such special 
potentials it is convenient to use Green's functions in coordinate
space when evaluating Feynman diagrams. Details of these calculations
are given in Ref.\cite{Gr03}. Here we content ourselves with listing
the results.

First, we consider two parallel plates in three spatial 
dimensions. The corresponding background in the sharp limit is
$\sigma_\parallel^\ast(\vec{x\,})=
\lambda\left[\delta(z-L)+\delta(z+L)\right]$. Then the $n^{\rm th}$ 
order contribution to the energy density per unit area is
$(\omega(q)=\sqrt{q^2+m^2})$
\begin{equation}
{\cal E}_\Lambda^{(n)}[\sigma^{\ast}_{\parallel}]=
\frac{-1}{2n}\hspace{-0.1cm}
\int^\Lambda \hspace{-0.15cm}\frac{d^{\, 3}q}{(2\pi)^3}
\hspace{-0.08cm}
\left[\frac{-\lambda}{2\omega(q)}\right]^n
\hspace{-0.1cm}
\left\{(1+{\rm e}^{-2\omega(q)L})^n+
(1-{\rm e}^{-2\omega(q)L})^n\right\}\,.
\label{enplates}
\end{equation}
Simple power counting shows that there are divergences at orders
$n=1,2$ and $3$, which are regulated by the sharp cutoff $\Lambda$.
The dependence on the separation is exponentially suppressed and
thus the divergences are separation independent. The orders $n\ge4$ 
are finite. Obviously this matches our expectations from  
the previous sections. Actually the Feynman series~(\ref{enplates})
can be summed:
\begin{equation}
{\cal E}_{\Lambda} [\sigma^{\ast}_{\parallel}]
=\int_m^{\sqrt{\Lambda^2+m^2}} \frac{dt}{(2\pi)^2}
t\,\sqrt{t^2-m^2}\ln \left[1+\frac{\lambda}{t}
+\frac{\lambda^2}{4t^2}\left(1-{\rm e}^{-4tL}\right)\right]\,.
\label{etotplates}
\end{equation}
The same result is found within the interface formalism\cite{Gr01}.
We can now compute a regularize pressure,
\begin{equation}
\mathcal{P}_{\Lambda}[\sigma_{\parallel}^{\ast}]
=-\frac{1}{2}\frac{\partial {\cal E}_{\Lambda}
[\sigma^{\ast}_{\parallel}] }{\partial L}=
-\frac{\lambda^2}{8\pi^2}
\int_m^{\sqrt{\Lambda^2+m^2}} dt\, \frac{\sqrt{t^2-m^2}\,{\rm
e}^{-4tL}} {1+\frac{\lambda}{t}+\frac{\lambda^2}{4t^2}
\left(1-{\rm e}^{-4tL}\right)}\,,
\label{prplates}
\end{equation}
which does not diverge as $\Lambda\to\infty$. Moreover, we can
take the strong limit $\lambda\to\infty$ and find the boundary
condition calculation  result\cite{Bo92} for the force between 
Dirichlet plates. This is expected, as the boundary condition 
approach is reliable for the force between rigid bodies.

We now repeat this exercise for the spherical shell, {\it i.e.}
$\sigma^{\ast}_{\circ}(\vec{x\,})=\lambda\delta(r-R)/4\pi R^{2}$ 
and find
\begin{eqnarray}
E_\Lambda^{(n)}(\sigma^{\ast}_{\circ})&=&
\frac{2}{n}\left(\frac{-\lambda}{4\pi R}\right)^n\,
\int_0^\Lambda \frac{d\omega}{2\pi}\,
I_n\left(R\sqrt{\omega^2+m^2}\right)\cr\cr
I_n(z)&=& \sum_{\ell}
\left(2\ell+1\right)\,
\left[I_{\ell+1/2}\left(z\right) K_{\ell+1/2}\left(z\right)\right]^n\,.
\label{ensphere}
\end{eqnarray}
The divergence structure of the momentum integral is dictated by the 
large $z$ behavior of the sum $I_n(z)$ which is shown for $n=2,3$ 
and $4$ in figure~\ref{fig1}. The sum $I_1(z)$ is actually divergent 
before integrating over the momentum. However, this is not a severe
problem because this contribution to the vacuum polarization energy
is completely canceled in the no--tadpole renormalization scheme.
\begin{figure}
\centerline{
\epsfig{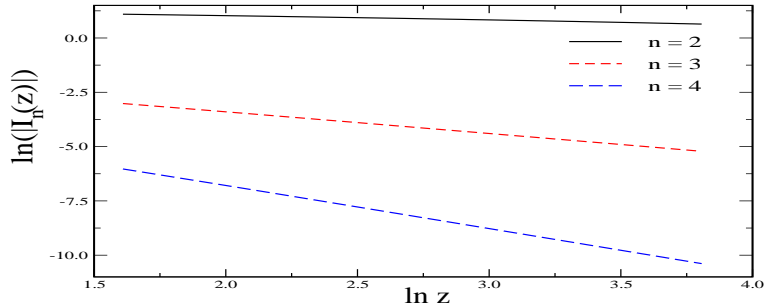}}
\caption{\label{fig2}Double logarithmic plot of the
orbital angular momentum sum in eq.~(\protect\ref{ensphere}).}
\end{figure}
We see from figure~\ref{fig1} that $I_n(z)\propto z^{-\alpha_n}$. Numerically
we find $\alpha_2=0.21$, $\alpha_3=1.00$ and $\alpha_4=1.99$. That is,
the second order contribution has a more severe singularity at large
cutoff than just logarithmically which was expected from the quantum 
divergence. This reflects the singular sharp limit that dwells on top of the 
quantum divergence. The third order contribution is logarithmically
divergent in support of the numerical result presented in the previous
section. Contributions of order four and higher are finite as 
$\Lambda\to\infty$. The important information is that the stress
$\partial E_\Lambda/\partial R$ is also ultraviolet divergent. That
is, the cutoff must be kept finite which is achieved by taking 
material properties into account. 

The expansion~(\ref{ensphere}) can formally be summed
leading to the same expression as found in the boundary condition
approach\cite{Mi02}. However, that is still a divergent quantity
when the cutoff is removed.

\section{Conclusions}

We have computed the Casimir energy of a Dirichlet boundary 
from singular limits of Casimir energies of (piecewise)
continuous background fields. For the latter, the Casimir energy
can be calculated using standard renormalization tools of quantum 
field theory. Although these energies are finite and unambiguous for 
any particular smooth background, the limit $\Delta\to0$, when the 
background becomes sharp and concentrated on the boundary, can still
diverge. For three space dimensions, the Casimir energy has 
$\frac{1}{\Delta}\ln\Delta$ divergences both for parallel plates and 
for the sphere. These divergences arise from large Fourier modes of 
the background field and should not be confused with the loop
divergences of Feynman diagrams: the former also appear in diagrams 
that have no quantum divergences. In particular, the third order 
diagram needs no renormalization but also diverges in the sharp limit, 
going like $\ln \Delta$.

This cutoff dependence may or may not enter into physically measurable
quantities.  The force between rigid bodies is never affected, so our
analysis does not alter standard results such as the force between
parallel plates.  However, it does render meaningless calculations of
stresses, such as the Casimir surface tension of the sphere.  Such
quantities are cutoff dependent, and cannot be defined
independently of the material properties that determine the cutoff.

We expect that these results generalize to the electromagnetic case.
Thus they invalidate Boyer's result\cite{Bo68} that the conducting sphere 
experiences a cutoff-independent, repulsive Casimir stress. For such a 
statement to hold, it is necessary to show that it can be obtained as the 
limit of an underlying smooth, renormalizable quantum field theory.  
Otherwise, the Boyer problem cannot be studied without reference to 
material properties, and the assertion that sphere has repulsive Casimir 
force is unwarranted.

\section*{Acknowledgments}

I would like to thank the organizers, especially Kim Milton, for
providing this interesting workshop. This work has been supported
in parts by the Deutsche Forschungsgemeinschaft (DFG) under 
contract We 1254/6--1.

\end{document}